\documentclass[review]{elsarticle}

\usepackage{hyperref}
\usepackage{amsmath}
\usepackage{amsthm} %problematic package
\usepackage{amssymb}
\usepackage{float}

\newtheorem{defi}{Definition}
\newtheorem{ex}{Example}

\newtheorem{lem}{Lemma}

\newtheorem{quest}{Question}

\newtheorem{thm}{Theorem}

\usepackage{color}

\providecommand{\modifq}[1]{{\color{black}#1}}
\providecommand{\modif}[1]{{\color{black}#1}}
\providecommand{\moddd}[1]{{\color{black}#1}}
\providecommand{\modq}[1]{{\color{black}#1}}
\providecommand{\modp}[1]{{\color{black}#1}}
\providecommand{\modr}[1]{{\color{black}#1}}
\providecommand{\mods}[1]{{\color{black}#1}}
\usepackage{psfrag}
\usepackage{tikz}
\usepackage{epstopdf}

\definecolor{lblue}{RGB}{200,200,200}
\definecolor{blue}{RGB}{50,50,50}
\definecolor{green}{RGB}{120, 120,120}
\definecolor{green2}{RGB}{10, 150, 10}

\newcommand{\vara}{10}
\newcommand{\varb}{5.6}
\newcommand{\varc}{5}
\newcommand{\vard}{1.2}
\newcommand{\sizerep}{1.5}
\newcommand{\decy}{3.5}
\newcommand{\decz}{\decy/2}

\newcommand{\ecv}{1.5} 
\newcommand{\ech}{1.3} 

\newcommand{\wi}{.6}

\journal{Systems \& Control Letters}

\begin{document}

\begin{frontmatter}

%Tight bounds for deciding convergence of consensus systems
\title{Tight Bounds for Deciding Convergence of Consensus Systems \tnoteref{mytitlenote}}
%\title{Tight Bounds for Consensus Systems Convergence\tnoteref{mytitlenote}}
\tnotetext[mytitlenote]{ R. M.  Jungers  is
an F.R.S.-FNRS research associate. This work  is also supported by the Belgian Network DYSCO, funded by the Belgian government and the Concerted Research Action (ARC) of the French Community of Belgium.}

%\author{P.-Y. Chevalier\corref{mycorrespondingauthor}, J. M. Hendrickx and R. M. Jungers\fnref{myfootnote}}
%\address{ICTEAM Institute, Universit\'e catholique de Louvain, 4 avenue Georges Lema\^itre,
%B-1348 Louvain-la-Neuve, Belgium}
%\fntext[myfootnote]{F.R.S.-FNRS Research Associate}

\author[add1]{P.-Y. Chevalier\corref{mycorrespondingauthor}}
%\author{P.-Y. Chevalier\corref{mycorrespondingauthor}}
\cortext[mycorrespondingauthor]{Corresponding author}
\ead{pierre-yves.chevalier@uclouvain.be}
\author[add1]{J. M. Hendrickx}
%\author{J. M. Hendrickx}
\ead{julien.hendrickx@uclouvain.be}
%\author[add1,add2]{R. M. Jungers}
\author[add1]{R. M. Jungers}
%\author{R. M. Jungers}
\ead{raphael.jungers@uclouvain.be}

\address[add1]{ICTEAM Institute, Universit\'e catholique de Louvain, 4 avenue Georges Lema\^itre, \\ B-1348 Louvain-la-Neuve, Belgium}
%\address[add2]{F.R.S.-FNRS Research Associate}
%\address[add3]{Department of Interests}


%\ead{pierre-yves.chevalier@uclouvain.be}

\begin{abstract}

We analyze the asymptotic convergence of all infinite products of matrices taken in a given finite set by looking only at \emph{finite} or \emph{periodic} products. 
It is known that when the matrices of the set have a common nonincreasing polyhedral norm, all infinite products converge to zero if and only if all infinite \emph{periodic} products with period\mods{s} smaller than a certain value 
converge to zero. Moreover,  bounds  on that value are available \cite{lagarias-finiteness}.

We provide a stronger bound that holds for both polyhedral norms and polyhedral seminorms. In the latter case, the matrix products do not necessarily converge to 0, but all trajectories of the associated system converge to a common invariant subspace.
We prove that our bound is tight for all seminorms.

Our work is motivated by problems in \emph{consensus systems}, where the matrices are \emph{stochastic} (nonnegative with rows summing to one), and hence always share a same common nonincreasing polyhedral seminorm. In that case, we also improve existing results.

\end{abstract}


\end{frontmatter}

\section{Introduction}

We consider the problem of determining the stability of matrix sets, that is, determining whether or not all infinite products of matrices from a given set converge to zero, or more generally to a common invariant subspace. 
This problem appears in several different situations in control engineering, computer science, and applied mathematics.  For instance, the stability  of matrix sets characterizes the stability of switching dynamical systems \cite{jungers_lncis1}, which have numerous application in control \cite{jungers_lncis1, Hernandez-Vargas, Shorten142}. 
Stability of matrix sets is instrumental in proving the continuity of certain wavelet functions \cite{jungers_lncis1, DaLa92}. Somewhat surprisingly, it also helped establishing the best known asymptotic bounds on the number of $\alpha$-power-free binary words of length $n,$ a central problem in combinatorics on words \cite{blondelcassaignejungers08, MoisionOrlitskySiegel01}.

Deciding the stability of a matrix set is notoriously difficult and the decidability of this problem is not known. 
The related problem of the existence of an infinite product whose norm diverges is undecidable \cite{BlTi2}.
However, it is possible to decide stability when the set has the \emph{finiteness property}, that is, when there is a bound $p$ such that the existence of an infinite nonconverging product\modq{\footnote{\modq{i.e., an infinite product that does not converge to zero\modr{.}}}} implies the existence of an infinite nonconverging \emph{periodic} product with period smaller than or equal to $p$. 
Indeed, %in this case,
%when such a bound $p$ exists, 
checking the stability of the set can be done by checking the stability of all products whose length is smaller than or equal to $p$. In this work, we look for the smallest valid bound $p$.

A similar question is particularly relevant in the context of consensus problems. These systems are models for groups of agents trying to agree on some common value by an iterative process. Each agent has a value $x_i$ which it updates by computing the weighted average of values of agents with which it can communicate. 
Consensus systems have attracted considerable attention due to their applications in control of vehicle formations \cite{Bamieh}, flocking \cite{Jad_tac, bhot05} or distributed sensing \cite{boyd_consensus, Olfati-Saber}.
They typically have \modq{time-}varying communication networks due to e.g. communication failures, or to the movements of the agents. This leads to systems whose (linear) dynamics may switch at each time-step.
\moddd{When a set of possible linear dynamics is known, one fundamental question is whether the system converges for any switching sequence \cite{bl-ol}.}

Consensus systems can be modeled by discrete-time linear switching systems, $x(t+1) = A_t x(t)$, where the transition matrices $A_t$ are \emph{stochastic} (nonnegative matrices whose rows sum to 1) because the agents always compute weighted averages.   \moddd{In this case, the products certainly do not converge to zero, since products of stochastic matrices remain stochastic. The central question is whether the agents asymptotically converge to the same value.} Deciding whether a consensus system converges for any sequence of transition matrices and any initial condition corresponds  to determining whether all left-infinite products of matrices taken from a set converge to a rank one matrix. Indeed, a stochastic matrix is rank one if and only if all its rows are the same, and this situation corresponds to consensus. % of the form $\textbf{1}y^\top$, where $y \in \mathbb{R}^n, \; y^\top \textbf{1} = 1$.
This particularization to stochastic matrices has other applications, including  inhomogeneous Markov chains, and probabilistic automata  \cite{Paz71}. 

\mods{Stochastic matrices share a \emph{nonincreasing polyhedral seminorm} and this property provides important information on the asymptotic convergence of products of these matrices.
%It turns out that the existence of an nonincreasing polyhedral seminorm provides strong information on the asymptotic convergence of products of these matrices.\\
Indeed, for sets of matrices sharing a common nonincreasing polyhedral seminorm, a bound $p$ as discussed above is available.}
%For sets of matrices with a common nonincreasing polyhedral \emph{norm}, there exists a bound $p$ such that the existence of an infinite nonconverging product implies the existence of an infinite nonconverging periodic product with period at most $p$. 
This was first established by Lagarias and Wang \cite{lagarias-finiteness}. The authors also give an explicit value for $p$ (namely half the number of faces of the unit ball of the norm). \moddd{ This result can easily be extended from norms to seminorms and we do so in the proof of Theorem \ref{main}\mods{.}\\}
The case of stochastic matrices has been analyzed earlier in the context of inhomogeneous Markov chains \cite{DaLa92, Paz71, Wolf63, AnTi77} and later in the context of consensus systems \cite{bl-ol}. 
A finiteness result has been known since Paz \cite{Paz71}, who proved that all left-infinite products converge to a rank one matrix  if and only if a certain condition on all products of length $B = \frac{1}{2} (3^n - 2^{n+1} + 1)$ is satisfied. %This bound has been rediscovered in the context of consensus \cite{bl-ol}. 
%This result can be interpreted as a bound $p$ such as described above for the particular case of stochastic matrices.
 In our recent paper \cite{CHJ12}, we showed that this bound can be derived from \mods{a generalization of} the result of Lagarias and Wang applied to \modq{a particular seminorm.} %the seminorm $\|.\|_\mathcal{P}$ described above.

\subsection*{Our Contribution}
In this \mods{article}, we consider a general problem that includes these particular cases: we study matrix sets for which there exists a polyhedral seminorm which is nonincreasing for all matrices in the given set, and we wonder whether long products of these matrices are asymptotically contractive.  We improve all the bounds previously known in the particular cases, and prove that our bound is tight. 
%More precisely, we answer the following question:
%\begin{quest}Let $\|.\|$ be a polyhedral seminorm; what is the smallest $p$ such that for any set $\Sigma$ for which $\|.\|$ is nonincreasing, the existence of an infinite noncontracting product implies the existence of an infinite \emph{periodic} noncontracting product with period smaller than or equal to $k?$
%\label{question}
%\end{quest}
%\vspace{-.6cm}
Our analysis relies on the fact that the convergence of the dynamical system \mods{can be} encapsulated in a \emph{discrete} representation by a dynamical system on the face lattice of the polyhedral (semi)norm. \modq{Our results then rely} on a careful study of the combinatorial structure of the trajectories in this discrete structure.

The improvement over the previously known bound depends on the seminorm. In the case of stochastic matrices, the improvement is a multiplicative factor  of about $ \frac{3}{2 \sqrt{\pi n}}$.

\section{Problem Setting}

\moddd{Let $\Sigma = \{A_1, \dots, A_m\}$ be a set of matrices and $\sigma$ an infinite sequence of indices. }
We say that the product $ \dots A_{\sigma(2)} A_{\sigma(1)}$ is \emph{periodic} if the sequence $\sigma$ is periodic. We recall that a \emph{seminorm} on $\mathbb{R}^n$ is an application $\|.\|$ with the following properties:
\begin{itemize}
\item $\forall x \in \mathbb{R}^n, \; a \in \mathbb{R}, \;\|ax\| = |a| \|x\|$
\item $\forall x, y \in \mathbb{R}^n, \; \|x + y \| \leq \|x\| + \|y\|$.
\end{itemize}
 We call a \emph{polyhedral seminorm} a seminorm whose unit ball is a \emph{polyhedron}, that is, a set that can be defined by a finite set of linear inequalities $$\{x : ||x|| \leq 1\} = \{x : \forall i, \; b_i^\top x \leq c_i \}.$$
We say that a seminorm $\|.\|$ is \emph{nonincreasing} with respect to a matrix $A$ if $$\forall x \in \mathbb{R}^n, \; \|Ax\| \leq \|x\|.$$
Geometrically, this correspond\mods{s} to its unit ball being invariant  $$A \{x : ||x|| \leq 1\} \subseteq \{x : ||x|| \leq 1\}.$$ We say that a seminorm is nonincreasing with respect to a set $\Sigma$ of matrices if it is nonincreasing with respect to each of the matrices in $\Sigma$.
 We say that a matrix $A$  \emph{contracts} a seminorm $\|.\|$ if $\forall x \in \mathbb{R}^n, \; \|Ax\| < \|x\|$. %We say that an infinite product $\dots A_{\sigma(2)} A_{\sigma(1)}$ does not contract a seminorm $\|.\|$ if  $$\forall t \geq 0, A_{\sigma(t)} \dots A_{\sigma(2)} A_{\sigma(1)} \mathcal{B}(\|.\|, 1, 0) \not\subset \mathcal{B}(\|.\|, 1, 0).$$}
We say that an infinite product $\dots A_{\sigma(2)} A_{\sigma(1)}$ contracts a seminorm $\|.\|$ if there is a $t$ such that $$A_{\sigma(t)} \dots A_{\sigma(2)} A_{\sigma(1)} \{x : ||x|| \leq 1\} \subset \text{int}(\{x : ||x|| \leq 1\}).$$
\moddd{One can easily verify that if there is a $p$ such that all products of length $p$ of matrices in $\Sigma$ contract a seminorm $\|.\|$,} then all trajectories $x(t)$ of the corresponding switching system $x(t+1) = A_{\sigma(t)} x(t)$ asymptotically approach the set $\{x:\|x\| = 0\}$, and that their distance to that set decays exponentially \modq{as $t$ increases}. In particular, if $\|.\|$ is a norm, $x$ converges exponentially to 0. \modq{In addition,} if $\|.\|$ is the seminorm $\|x\|_{\mathcal{P}}  = \frac{1}{2}(\max_i x_i - \min_i x_i)$ -- a seminorm that is nonincreasing for stochastic matrices -- then $x$ approaches the \emph{consensus space} $\{\alpha \mathbf{1}\}$. We have proved in previous work \cite{CHJ12} that each trajectory actually converges in that case to a specific \moddd{(but possibly different)} point in that set, as opposed to just approaching the set. For these reasons, we will investigate contraction of seminorms, keeping in mind that this question is intimately related to that of convergence.

%\addtocounter{quest}{-1}
%\begin{quest}Let $\|.\|$ be a seminorm, what is the smallest $p$ such that for any set $\Sigma$ for which $\|.\|$ is nonincreasing, the existence of an infinite noncontracting product implies the existence of an infinite periodic noncontracting product with period smaller than or equal to $p$?
%\end{quest}
\begin{quest}Let $\|.\|$ be a polyhedral seminorm \modq{in $\mathbb{R}^n$ for some fixed $n$}; what is the smallest $p$ such that for any set $\Sigma$ for which $\|.\|$ is nonincreasing, the existence of an infinite noncontracting product implies the existence of an infinite \emph{periodic} noncontracting product with period smaller than or equal to $p?$
\label{question}
\end{quest}

\vspace{-.8cm}

\section{The General Case}

\modq{In this section, we answer Question \ref{question}.}
We start by recalling some definitions \modq{(see \cite{ziegler} for more details).}  %As regards poset and polyhedron terminology, we use the definitions of \cite{ziegler}.
A \emph{partially ordered set} \mods{or \emph{poset}} is a set $P$ with a binary relation $\preceq$ that is transitive, antisymmetric and reflexive. We also note $x \prec y$ \modq{for} the relation $x \preceq y \text{ and } x \neq y$. %An \emph{antichain} is a set $X \subseteq P$ of elements that are not comparable: $$\forall x, y \in X, \; x \nprec y \text{ and } y \nprec x.$$
A poset $(P, \preceq)$ is called \emph{graded} if it can be equipped with a rank function $r: P \mapsto \mathbb{N}$ such that $x \preceq y \Rightarrow r(x) \leq r(y) $ and $\left( y \prec x  \text{ and } \nexists z, \; y \prec z \prec x \right) \Rightarrow r(x) = r(y) + 1.$ \mods{The set of all elements of a given rank is called a rank level.% is a set$\{x \in P \; | \; r(x) = c\}$.
} %if and only if $x$ covers $y$, that is, $y \preceq x$ and $\nexists z, \; y \preceq z \preceq x$
A poset is called a \emph{lattice} if any pair of elements has a unique infimum and a unique supremum.

%\comjh{below: explain what affinely independent means}
%
%$u_0, u_1, \dots, u_d$
%
%$u_1 - u_0, u_2 - u_0, \dots, u_d - u_0$

\moddd{Intuitively, a \modq{\emph{face}} is the generalization of a vertex \modq{(or an edge, or a facet)} to an arbitrary dimension. The formal definition is the following.}
%\moddd{We call a \emph{face} of a polyhedron a vertex,  an edge, a facet or the generalization of these to an arbitrary dimension.}

\begin{defi}[Faces of a Polyhedron] A nonempty subset $F$ of an $n$-dimensional polyhedron $\mathcal{Q}$ is called a \emph{face} or \emph{closed face} if \modq{one of the following holds: 
\begin{itemize}
\item $F = \mathcal{Q}$,
\item $F = \varnothing$ 
\item or $F$ can be represented as $F = \mathcal{Q} \cap \{ x : b^\top x = c\}$ where $b \in \mathbb{R}^n$, 
$c \in \mathbb{R}$ are such that $$\forall x \in \mathcal{Q}, \; b^\top x \leq c.$$
\end{itemize}}
If the face contains exactly $d+1$ affinely independent points\footnote{\modq{The points $u_0, u_1, \dots, u_d$ are called affinely independent if  $u_1 - u_0, u_2 - u_0, \dots$ are linearly independent.}},
  we call $d$ the \emph{dimension} of the face. 
A \emph{proper face} is a face that is neither the polyhedron itself nor the empty face.
An \emph{open face} is the relative interior of a face. 
%We call a \emph{face} of Polyhedron $\mathcal{Q}$
\moddd{Finally, a \emph{facet} is a face of dimension $n-1$.}
\end{defi}
It is well known that faces of any dimension are intersections of facets and their number is therefore finite.
It is also known that any polyhedron decomposes into a disjoint union of open faces.

\modp{We use the term \mods{\emph{double-face} to denote the} set $F \cup -F$, for some proper face $F$. %\mods{We say that a double-face $F \cup -F$ has dimension $d$ if $F$ has dimension $d$.}
 A \mods{double-face} is called open \modr{if} the face $F$ is open, and closed otherwise.}

%\begin{defi}[Face Lattice] Given a polyhedron $\mathcal{Q}$, we call \emph{face lattice} the poset $(P, \subseteq)$ where $P$ is the set of (closed) faces of $\mathcal{Q}$  and $F_1 \subseteq F_2$ is the inclusion relation. %of Closed faces ordered by inclusion. $$F_1 \preceq F_2 \Leftrightarrow F_1 \subseteq F_2.$$
%In the case of polytopes, it is well-known that this poset is a graded lattice. This is also the case for unbounded centrally symmetric polyhedra. A rank function is given by the dimension of the faces and $d_{\min} - 1$ for the empty face, where $d_{\min}$ is the lowest dimension of faces of $\mathcal{Q}$.
%\end{defi}

\mods{
\begin{defi}[\mods{lattice of double-faces}] Given a centrally symmetric  polyhedron $\mathcal{Q}$ (i.e., a polyhedron $\mathcal{Q} = - \mathcal{Q}$), we call \emph{lattice of \mods{double-faces}}  the poset $(P, \subseteq)$ where $\subseteq$ is the inclusion relation and $P$ is a set whose members are 
\begin{itemize}
\item \mods{double-faces}  of $\mathcal{Q}$ ($r=$ dimension of the face)
\item $\mathcal{Q}$ ($r = n$)
\item $\varnothing$ ($r=d_{\min} - 1$, where $d_{\min}$ is the lowest dimension of faces of $\mathcal{Q}$).
\end{itemize}
   %of Closed faces ordered by inclusion. $$F_1 \preceq F_2 \Leftrightarrow F_1 \subseteq F_2.$$
%In the case of polytopes, it is well-known that this poset is a graded lattice. This is also the case for unbounded centrally symmetric polyhedra. A rank function is given by the dimension of the faces and $d_{\min} - 1$ for the empty face, where $d_{\min}$ is the lowest dimension of faces of $\mathcal{Q}$.
It can be verified that this poset is a lattice and that it is graded; a rank function is given between brackets.
%\mods{
%\begin{itemize}
%\item$d_{\min} - 1$ for the empty face, where $d_{\min}$ is the lowest dimension of faces of $\mathcal{Q}$
%\item the dimension of the double-face for  face.
%\end{itemize}}
\end{defi}
}

\begin{defi}[Antichain] Let $(P, \preceq)$ be a poset. An \emph{antichain} is a subset $S \subseteq P$ whose elements are not comparable: $$\forall S_1, S_2 \in S, \;\; S_1\not\preceq S_2.$$
\end{defi}
\modq{For instance, a set of \mods{double-faces} that are not included in one \mods{another} form an antichain in the \mods{lattice of double-faces}.}

\begin{ex}\modq{
 The unit ball of the seminorm $\|x\|_\mathcal{P} = \frac{1}{2}(\max_i x_i - \min_i x_i)$ in dimension 3 is represented in Figure \ref{P}. We will study this seminorm and \modr{its} relation to stochastic matrices in detail in the next section.}
\modq{
The \mods{lattice of double-faces} of this unit ball and its largest antichain are represented in Figure \ref{lattice}.}
\begin{figure}[h]
\centering
\usetikzlibrary{calc}

%\begin{tikzpicture}[scale=1.2,looseness=1,auto,line width=.5mm,shorten >=3pt, x= {(-0.353cm,-0.353cm)}, z={(0cm,1cm)}, y={(1cm,0cm)}]
\begin{tikzpicture}[scale=.6,looseness=1,auto,line width=.5mm, x= {(-0.6095cm,-0.4265cm)}, z={(0cm,1cm)}, y={(0.933cm,-0.25cm)}]
%\begin{tikzpicture}[scale=1.2,looseness=1,auto,line width=.5mm,shorten >=3pt, x= {(-0.866cm,-0.5cm)}, z={(0cm,1cm)}, y={(0.866cm,-0.5cm)}]
\begin{scope}[every node/.style={font=\small\itshape}]
\tikzstyle{every node}=[draw=none];

		% coordonnées du repere
		\coordinate (12) at (\varb, 0, 0);
		\coordinate (13) at (0, \varb, 0);
		\coordinate (14) at (0, 0, \varb);
		\coordinate (15) at (1.5*\vard, 0, 0);
		\coordinate (or) at (0, 0, 0);
		
		% dessin du repere
		%\draw[->, line width=.5mm] (or) -- (12);
		\draw[->, line width=.5mm] (or) -- (13);
		\draw[->, line width=.5mm] (or) -- (14);
			
		% coordonnees du polyèdre
        \coordinate (0) at ( \vard - \varc,  -\vard/2 - \varc, -\vard/2 - \varc);
        \coordinate (1) at ( \vard/2 - \varc,  -\vard - \varc, \vard/2 - \varc);
        \coordinate (2) at ( -\vard/2 - \varc,  -\vard/2 - \varc, \vard - \varc);
        \coordinate (3) at ( -\vard - \varc,  \vard/2 - \varc, \vard/2 - \varc);
        \coordinate (4) at ( -\vard/2 - \varc,  \vard - \varc, -\vard/2 - \varc);
        \coordinate (5) at ( \vard/2 - \varc, \vard/2 - \varc, -\vard - \varc);
        
        \coordinate (6)  at ( \vard + \vara,  -\vard/2 + \vara, -\vard/2 + \vara);
        \coordinate (7) at ( \vard/2 + \vara,  -\vard + \vara, \vard/2 + \vara);
        \coordinate (8)  at ( -\vard/2 + \vara,  -\vard/2 + \vara, \vard + \vara);
        \coordinate (9)  at ( -\vard + \vara,  \vard/2 + \vara, \vard/2 + \vara);
        \coordinate (10)  at ( -\vard/2 + \vara,  \vard + \vara, -\vard/2 + \vara);
        \coordinate (11) at ( \vard/2 + \vara,  \vard/2 + \vara, -\vard + \vara);
        
        \coordinate (cons) at (1.6*\vara, 1.6*\vara, 1.6*\vara);

		 % faces arrieres
		 \draw[fill=lblue   ,opacity=1] (2) -- (1) -- (7) -- (8);
		 \draw[fill=lblue   ,opacity=1] (2) -- (3) -- (9) -- (8);
		 \draw[fill=lblue   ,opacity=1] (4) -- (3) -- (9) -- (10);	
		 
		 % aretes arrieres	 
		 \draw[-, line width=.5mm] (2) -- (8);
		 \draw[-, line width=.5mm] (3) -- (9);
		 % vecteur consensus
		 \draw[->, line width = .5mm, green] (or) -- (cons);
		 %faces avant
		 \draw[fill=lblue   ,opacity=1] (4) -- (5) -- (11) -- (10);
		 \draw[fill=lblue   ,opacity=1] (0) -- (5) -- (11) -- (6);
		 \draw[fill=lblue   ,opacity=1] (0) -- (1) -- (7) -- (6);

         % arêtes du polytope
         \draw[-, line width=.5mm] (0) -- (6);
		 \draw[-, line width=.5mm] (1) -- (7);
		 \draw[-, line width=.5mm] (4) -- (10);
		 \draw[-, line width=.5mm] (5) -- (11);
		 
		 % axe qui sort du polytope
		 \draw[->, line width=.5mm] (15) -- (12);

		%\draw[dashed, ->, line width=.5mm] (or) -- (12);
		%\draw[dashed, ->, line width=.5mm] (or) -- (13);
		%\draw[dashed, ->, line width=.5mm] (or) -- (14);
		\draw[dashed, ->, line width = .5mm, green] (or) -- (cons);
		%\draw[dashed, -] (2) -- (8);
		%\draw[dashed, -] (3) -- (9);
		%\draw[dashed, -] (1) -- (2);
		%\draw[dashed, -] (3) -- (2);
		%\draw[dashed, -] (3) -- (4);

		% texte (placement manuel)
		\node[text width=1cm] at (0,3.2,2) {$f_1$};
		\node[text width=1cm] at (0,4,5) {$f_1$};

%		\node[text width=1cm] at (0,0) {$f_2$};
%		\node[text width=1cm] at (0,0) {$f_2$};

		\node[text width=1cm] at (0,5,4) {$f_3$};
		\node[text width=1cm] at (8,7,7.5) {$f_3$};

		\node[text width=1cm] at (.3,.55,0) {$e_3$};
		\node[text width=1cm] at (13,14.3,13.2) {$e_3$};

\end{scope}
\end{tikzpicture}
\caption{The polyhedron $\mathcal{P}$ for $n=3$. The gray arrow indicates the direction $a \textbf{1}.$ The polyhedron has 6 facets\mods{, one for each constraint} of the form $\frac{1}{2}(x_i - x_j) \leq 1.$ The sets $f_1$, $f_3$ and $e_3$ are \mods{double-faces}.}
\label{P}
%ordered pair} of the three variables $x_1, x_2, x_3$ and, therefore, 6
\end{figure}
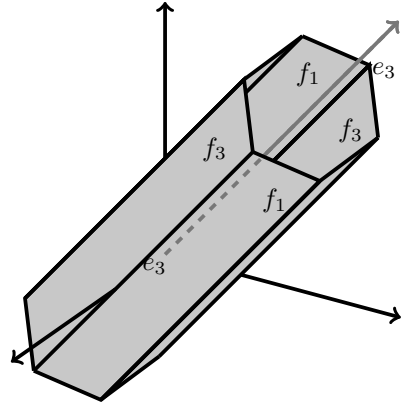
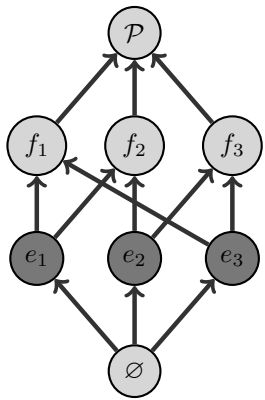
\begin{figure}[h!]
\centering
\usetikzlibrary{arrows}

%\begin{tikzpicture}[->,>=stealth',shorten >=1pt,auto,node distance=2cm,
%  thick,main node/.style={circle,fill=blue!20,draw,font=\sffamily\Large\bfseries}]
%\begin{tikzpicture}[->,shorten >=1pt,auto,node distance=2cm,
%  thick,main node/.style={circle,fill=blue!20,draw}]
\begin{tikzpicture}[->,auto,node distance=2cm,
  thick,main node/.style={circle,fill=blue!20,draw}]
  %\begin{tikzpicture}[->,shorten >=1pt,auto,node distance=2cm,
  %thick,main node/.style={circle,fill=blue!20,draw,font=\sffamily\Large\bfseries}]

  \node[main node] (0)   at (0, 0) {$\varnothing$};

% edges
  \node[main node, fill=green] (11) at (-\ech, \ecv) {$e_1$};
  \node[main node, fill=green] (12) at (0, \ecv) {$e_2$};
  \node[main node, fill=green] (13) at (\ech, \ecv) {$e_3$};

% faces
  \node[main node] (21) at (-\ech, 2*\ecv) {$f_1$};
  \node[main node] (22) at (0, 2*\ecv) {$f_2$};
  \node[main node] (23) at (\ech, 2*\ecv) {$f_3$};

  \node[main node] (3)   at (0, 3*\ecv) {$\mathcal{P}$};

   \path[every node/.style={font=\sffamily\small}]
% arêtes du bas
   (0) edge [draw=blue, line width=\wi mm] node [left] {} (11)
   (0) edge [draw=blue, line width=\wi mm] node [left] {} (12)
   (0) edge [draw=blue, line width=\wi mm] node [left] {} (13)

% arêtes du milieu
   (11) edge [draw=blue, line width=\wi mm] node [left] {} (21)
   (12) edge [draw=blue, line width=\wi mm] node [left] {} (22)
   (13) edge [draw=blue, line width=\wi mm] node [left] {} (23)

   (11) edge [draw=blue, line width=\wi mm] node [left] {} (22)
   (12) edge [draw=blue, line width=\wi mm] node [left] {} (23)
   (13) edge [draw=blue, line width=\wi mm] node [left] {} (21)

% arêtes du haut
   (21) edge [draw=blue, line width=\wi mm] node [left] {} (3)
   (22) edge [draw=blue, line width=\wi mm] node [left] {} (3)
   (23) edge [draw=blue, line width=\wi mm] node [left] {} (3)

;
\end{tikzpicture}
\caption{The \mods{lattice of double-faces} of the polyhedron $\mathcal{P}$ for $n=3$. The elements $f_1$, $f_2$ and $f_3$ represent the three pairs of opposite facets while $e_1$, $e_2$ and $e_3$ represent the three pairs of opposite edges. In dark gray, a largest antichain in this lattice.}
\label{lattice}
\end{figure}
\end{ex}

\begin{defi}[Width of a Poset] We call the width $W(P)$ of a poset $P$ the number of elements of the largest antichain of $P$. %For graded posets the width is equal to the largest rank level:
% $$W(P) = \max_k \#(\{x \; | \; rank(x) = k\}).$$
 We also write $W(\mathcal{Q})$ for the width of the \modp{\mods{lattice of double-faces} of a given centrally symmetric} polyhedron $\mathcal{Q}$.
\end{defi}

The following lemma by Lagarias and Wang allows abstracting Question 1 as a combinatorial problem, as it shows that matrices in $\Sigma$ can be completely \modq{abstracted} (for our purpose) as functions \mods{mapping} each face of the invariant polyhedron \mods{into} another one.

\begin{lem} Let $\Sigma$ be a finite set of matrices having a common invariant polyhedron $\mathcal{Q}$. 
Then, for any $A \in \Sigma$ and any \modp{\mods{double-face}}  $O_1$ of $\mathcal{Q}  $, 
 there exists exactly one \modp{\mods{double-face}} $O_2$ (possibly $\text{int}(\mathcal{Q})$) such that $$A O_1 \; \subseteq \; O_2.$$
\modp{
\begin{proof} The result is established in \cite[Claim in the proof of Theorem 4.1]{lagarias-finiteness} for faces instead of \mods{double-faces}. \modr{It is clear that  the  open faces $O_1$, $O_2$ satisfy $A O_1 \; \subseteq \; O_2$ if and only if the open \mods{double-faces} $O_1 \cup - O_1$ and $O_2 \cup - O_2$ satisfy $A (O_1 \cup - O_1) \; \subseteq \; (O_2 \cup - O_2).$  The result therefore extends to \mods{double-faces}.} %The extension to \mods{double-faces} is straightforward.
\end{proof}
}
\label{LW}
\end{lem}
%\vspace{-0cm}

The next theorem is an improvement of \cite[Theorem 4.1]{lagarias-finiteness}. We extend it to seminorms and we provide a stronger bound\modq{.} %: half the size of the largest antichain instead of half the total number of faces.
\begin{thm} Let $\Sigma$ be a set of matrices and let $\|.\|$ be a polyhedral seminorm that is nonincreasing for $\Sigma$. If there is a left-infinite product of matrices from $\Sigma$ that does not contract $\|.\|$, there is one that is periodic with a period $p$ not larger than  \modp{$$p^* = W(\mathcal{B}), \text{ with } \mathcal{B} = \{x : ||x|| \leq 1\}.$$}% $$p^* = \frac{W(\mathcal{B})}{2}, \text{ with } \mathcal{B} = \{x : ||x|| \leq 1\}.$$ 
\begin{proof} 

We first prove that $p$ is finite. \modr{Suppose} there \modq{exists an} infinite noncontracting product $\dots A_{\sigma(2)}  A_{\sigma(1)}$ and therefore a point $x_0$ such that $$\forall i, \; A_{\sigma(i)} \dots A_{\sigma(1)} x_0 \notin \text{int}(\mathcal{B}).$$
Since the number of faces is finite, there is  \modp{an open \mods{double-face} $O$} and indices $i < j$ such that $$A_{\sigma(i)} \dots A_{\sigma(1)} x_0 \in O \text{ and } A_{\sigma(j)} \dots A_{\sigma(1)} x_0 \in O.$$
By Lemma \ref{LW}, we have $$A_{\sigma(j)} \dots A_{\sigma(i+1)} O \subseteq O.$$ Therefore, the infinite power of $A_{\sigma(j)} \dots A_{\sigma(i+1)}$ is an infinite \emph{periodic} noncontracting product, proving that the theorem is true for some finite period $p = j-i$ \modp{smaller than the number of \mods{double-faces}.} % \leq \# faces(\mathcal{B})$.  

We now prove the full theorem.
Let \modr{$P$ be such that \modq{$\dots PPP$} is an} infinite noncontracting product with the smallest period $p$ and $$P = A_{\sigma(\modq{p})} \dots A_{\sigma(1)}.$$ Let $O_1$ be \modp{a \mods{double-face}} %an open face of $\mathcal{B}$ 
such that $$\forall t \geq 0, \; (P)^t O_1 \not\subseteq \text{int}(\mathcal{B})$$ (such a face exists due to Lemma \ref{LW} and the fact that $\dots PPP$ is noncontracting), let $O_2$ be the \modp{\mods{double-face}} containing  $A_{\sigma(1)}O_1$ (by Lemma \ref{LW}, there is exactly one such \modp{\mods{double-face}}), $O_3$ containing $A_{\sigma(2)}A_{\sigma(1)}O_1$ up to $O_{p}$ containing $A_{\sigma(p-1)} \dots A_{\sigma(1)}O_1$. %Let $F_i = \text{cl}(O_i)$. 
Let also $F_1 = \text{cl}(O_1)$, $\dots,$ $F_{p} = \text{cl}(O_{p})$.

%\modq{We will prove that the cardinality $p$ of the set $\{F_1, \dots, F_{p}\}$ is at most half the size of the largest antichain in the face lattice of $\mathcal{B}.$ For that, we will prove that for $i\neq j$, then $F_i \not\subseteq F_j$ and $F_i \not\subseteq -F_j$. It then follows directly that $\{F_1, \dots, F_{p}, -F_1, \dots, F_{p}\}$ is an antichain and therefore not larger than the largest antichain, hence $p \leq p^*$.}
\modp{We now prove that $\{F_1, \dots, F_p\}$ in an antichain in the \mods{lattice of double-faces}. 
%We prove that $\{F_1, \dots, F_{p}\}$ is an antichain in the face lattice of $\mathcal{B}$. %is no $i \neq j$ such that $O_i \subseteq \text{cl}(O_j)$. }
Suppose, to obtain a contradiction, that for \modif{some $i, j$ with $i > j$, }
% option 1
%\modq{$F_i \subseteq F_j$ or $F_i \subseteq -F_j$.} 
% option 2
%$\modifq{F_i} \subseteq  \modq{(F_j \cup -F_j)}$. Then %by Lemma \ref{LW} 
% option 3
\modr{$F_i \subseteq F_j$. Then,}
$$A_{\sigma(i-1)} \dots A_{\sigma(j)} F_j \subseteq F_i \mods{\subseteq F_j},$$   
%$$A_{\sigma(i-1)} \dots A_{\sigma(j)} F_j \subseteq F_i \subseteq  F_j,$$   
%$$A_{\sigma(i-1)} \dots A_{\sigma(j)} F_j \subseteq F_i \subseteq   \modq{(F_j \cup -F_j)},$$   
%$$A_{\sigma(i-1)} \dots A_{\sigma(j)}\text{cl}(O_j) = \text{cl}(A_{\sigma(i-1)} \dots A_{\sigma(j)}O_j) \subseteq  \text{cl}(O_j)$$
 and thus
$$\forall t \geq 0, \; (A_{\sigma(i-1)} \dots A_{\sigma(j)})^{t} F_j \subseteq F_j.$$ }
This contradicts the assumption that $\dots PPP$ is the infinite periodic noncontracting product with the smallest period. 
%\modq{Similarly, an infinite noncontracting product with a period shorter than the length of $P$ could be constructed if $F_i \subseteq F_j$ for some $i>j$ or $F_i \subseteq F_j$ or $F_i \subseteq -F_j$ for some $i<j$.}
\modp{Similarly, if for some $i, j$ with $i < j$, $F_i \subseteq  F_j$, then $$\forall t \geq 0, \; (A_{\sigma(i-1)} \dots A_{\sigma(1)} A_{\sigma(p)} \dots A_{\sigma(j)})^t F_j \subseteq  F_j,$$ and again we have a contradiction.}
%Similarly, if for some $i < j$, $F_i \subseteq  F_j$\modq{, or $F_i \subseteq  -F_j$, then $A_{\sigma(i-1)} \dots A_{\sigma(1)} A_{\sigma(p)} \dots A_{\sigma(j)}$ would be }%, then $$\forall t \geq 0, \; (A_{\sigma(i-1)} \dots A_{\sigma(1)} A_{\sigma(p)} \dots A_{\sigma(j)})^t F_j \subseteq  F_j,$$ and again we would have a contradiction.
%Therefore, there is no $i \neq j$ such that $F_i \subseteq F_j$, and the set of faces $\{F_1, \dots, F_{p}\}$ is an antichain. 

%We now prove that $\{O_1, \dots, O_{p}\}$ does not contain opposite faces: if it contains opposite faces $O_i = - O_j$, then the product $P_2 = A_{\sigma(i-1)} \dots A_{\sigma(j)}$ is shorter than $P$ and its powers are noncontracting \moddd{because $P_2^2 O_j \subseteq O_j$}, again contradicting our assumption. %\modif{that $PP\dots$ is the noncontracting infinite product with the smallest period.}

%The set of faces $\{F_1, \dots, F_{p}\}$ is thus an antichain and does not contain opposite faces. Since each proper face has an opposite face, every antichain that does not contain opposite faces has at most $p^* = \frac{W(\mathcal{B})}{2}$ elements.
%Therefore, $p \leq p^*$, which concludes the proof.
\end{proof}
\label{main}
\end{thm}

The bound $p^*$ of Theorem \ref{main} cannot be decreased: it is tight for any polyhedron\mods{,} as we show next.
\begin{thm} Let $\|.\|$ be a polyhedral seminorm. There is a set $\Sigma$ for which $\|.\|$ is non increasing and such that 
\begin{itemize}
%\item $\|.\|$ is nonincreasing for $\Sigma$,
\item all infinite periodic products with period\mods{s} smaller than $p^*$ are contracting,
\item not all products are contracting.
\end{itemize}
\begin{proof}
%\textbf{Claim 2:} $p^* \geq \frac{W(\mathcal{B})}{2}$. 
\modif{Let again $\mathcal{B} = \{x : ||x|| \leq 1\}$. We construct a set of matrices such that the infinite noncontracting product that has the smallest period has a period equal to \modr{$p^*=W(\mathcal{B})$}.} Let $X = \{F_1, \dots, F_{p^*}\}$ be the largest antichain \modp{in the \mods{lattice of double-faces}} % that does not contain opposite faces % in the face lattice, let  
% $\{F_1, \dots, F_m\}$ be the largest subset of $X$ that does not contain opposite faces and let 
and let $O_1, \dots, O_{p^*}$ be the corresponding open \mods{double-faces}. %We have that $m \geq \frac{l}{2}$. % \geq \frac{W(\mathcal{Q})}{2}$.

By definition, each \modp{\mods{double-face} $F_i$ is the union of two opposite proper faces $G_i, -G_i$ and the proper face $G_i$ is the intersection of $\mathcal{B}$ with a hyperplane}
%proper face \modp{$G$} is the intersection of $\mathcal{B}$ with a hyperplane
$$\modp{G_i} = \mathcal{B} \cap \{x : b_i^\top x = c_i\}$$ such that $\mathcal{B}$ is in one halfspace defined by the hyperplane: %$$.$$ $b$, $c$ such that 
%$$\forall x \in \mathcal{Q}, \; b^\top x \leq c.$$
$$ \mathcal{B} \subseteq \{x : b_i^\top x \leq c_i\}.$$ 
 \mods{
We also have $c_i \neq 0$.
Indeed, if $c_i = 0$, then $\mathcal{B} \subseteq \{x : b_i^\top x \leq 0\}$ and because , $\mathcal{B}$ is the unit ball of a seminorm, $\mathcal{B} = -\mathcal{B}$, and $\mathcal{B} \subseteq \{x : - b_i^\top x \leq 0\}$ and this implies $\modp{G_i} = \mathcal{B} \cap \{x : b_i^\top x = 0\} = \mathcal{B}$ and \modp{$G_i$} is not a proper face. Therefore, $c_i \neq 0$ and we can scale $b_i$ and $c_i$ to have $\forall i, \; c_i = 1$. \modp{Finally, $F_i = G_i \cup -G_i = \mathcal{B} \cap \{x : b_i^\top x = \pm1\}$.}}

By taking any $v_i$ in the open \mods{double-face} $O_{(i \text{ mod } p^*) + 1}$ and defining 
$$A_i = v_{i} b_i^\top \text{ and } \Sigma = \{A_1, \dots , A_{p^*}\}, $$% = \{v_{1} b_1^\top, \dots , v_{m} b_m^\top\},$$
we have
\begin{equation} \begin{aligned} \forall i, \; A_i F_i &= A_i (\mathcal{B} \cap \{x : b_i^\top x = \modp{\pm} 1 \}) \\
&\subseteq A_i  \{x : b_i^\top x = \modp{\pm}1 \} \\
&= \{A_ix : b_i^\top x = \modp{\pm}1\}\\
&= \{v_i b_i^\top x : b_i^\top x = \modp{\pm}1\} \\
&= \{\modp{\pm}v_i\} \\
&\subseteq O_{(i \text{ mod } p^*) + 1}.\end{aligned}\label{AF}\end{equation} 
%\begin{equation} \begin{aligned} \forall i, \; A_i F_i &= A_i (\mathcal{B} \cap \{x : b_i^\top x = 1 \}) \\
%&\subseteq A_i  \{x : b_i^\top x = 1 \} \\
%&= \{A_ix : b_i^\top x = 1\}\\
%&= \{v_i b_i^\top x : b_i^\top x = 1\} \\
%&= \{v_i\} \\
%&\subseteq O_{(i \text{ mod } p^*) + 1}.\end{aligned}\label{AF}\end{equation} %which verifies (\ref{AF}). 
%Because $\mathcal{B}$ is centrally symmetric, \modq{$- F_i$ is a face and, morevover,} it is equal to $\{x : -b_i^\top x = 1 \}$. We then have
\modp{We have as well
\hspace{-1.2cm}
\begin{equation}\begin{aligned} \forall i, \; A_i (\mathcal{B} \backslash F_i) 
&= A_i (\mathcal{B} \cap \{x : -1 < b_i^\top x < 1 \}) \\
&\subseteq \{v_i b_i^\top x : -1 < b_i^\top x < 1\} \\
&= \{\lambda v_i : -1< \lambda <1\}\\
&\subseteq \{\lambda y : -1< \lambda <1, \; y \in \mathcal{B} \} \\
&= \text{int}(\mathcal{B}).
\end{aligned}\label{AQ}\end{equation}}
%\begin{equation}\begin{aligned} \forall i, \; A_i (\mathcal{B} \backslash (F_i \cup - F_i)) 
%&= A_i (\mathcal{B} \cap \{x : -1 < b_i^\top x < 1 \}) \\
%&\subseteq \{v_i b_i^\top x : -1 < b_i^\top x < 1\} \\
%&= \{\lambda v_i : -1< \lambda <1\}\\
%&\subseteq \{\lambda y : -1< \lambda <1, \; y \in \mathcal{B} \} \\
%&= \text{int}(\mathcal{B}).
%\end{aligned}\label{AQ}\end{equation}

By  (\ref{AF}) and (\ref{AQ}), for any $j \neq (i \text{ mod } p^*) + 1 $ and any subset $S$ of $\mathcal{B}$,
\begin{align*} A_j A_i S &\subseteq  A_j \left(\text{int}(\mathcal{B}) \cup O_{(i \text{ mod } p^*) + 1}\right) \\ &=  A_j \text{int}(\mathcal{B}) \cup A_j O_{(i \text{ mod } p^*) + 1} \subseteq \text{int}(\mathcal{B}).\end{align*}
%$$A_j A_i S \subseteq A_j \left(\text{int}(\mathcal{B}) \cup O_{(i \text{ mod } m) + 1}\right) = A_j \text{int}(\mathcal{B}) \cup A_j O_{(i \text{ mod } p^*) + 1} \subseteq \text{int}(\mathcal{B}).$$
Therefore, $$\dots A_{(h + 2 \text{ mod } p^*) + 1} A_{(h + 1 \text{ mod } p^*) + 1} A_{(h \text{ mod } p^*) + 1} A_h$$ is the only infinite noncontracting product starting with $A_h$. For any $h$, this product has a period of $p^*$ (because the matrices  $A_1, \dots, A_{p^*}$ are all different). %With $X$ being the largest antichain, we obtain $m = \frac{l}{2} =  \frac{W(\mathcal{B})}{2} = p^*$. 
We conclude that all infinite periodic products with period\mods{s} smaller than $m = p^*$ are contracting and the theorem is proven.
%That is, the noncontracting infinite product with the smallest period has a period of $m$. With $X$ being the largest antichain, we obtain $m = \frac{l}{2} =  \frac{W(\mathcal{B})}{2}$ and we conclude that $p^* \geq \frac{W(\mathcal{B})}{2}$.
\end{proof}
\label{tightness_main}
\end{thm}

Giving an explicit value to the size of the largest antichain may prove difficult in some cases.
%is not always an easy task. 
However, since a set of \mods{double-faces} of same dimension always constitute an antichain, \modq{the largest antichain has at least $\max_i f_i$ elements, and} we have the following lower bound 
%a lower bound on the size of the largest antichain is given by
%$$p^* = \frac{W(\mathcal{B})}{2}  \geq \frac{\max_i f_i}{2},$$
\modp{$$p^* = W(\mathcal{B})  \geq \mods{\max_i f_i},$$}
where $f_i$ is the number of faces of dimension $i$.
\mods{If the equality holds, the exact value of $p^*$ can be known. This is the case when} the \modp{\mods{lattice of double-faces}} of $Q$ has the \emph{Sperner property}\modq{:}

\begin{defi}[Sperner Property \cite{Sperner_theory}] A graded poset is said to have the Sperner property \mods{if the largest antichain is equal to the largest rank level.} %if no antichain in it is larger than the width of the poset.  
\end{defi}

%Not all face lattices have the Sperner property. % In appendix A, Example 1 presents seminorm that has a unit ball whose face lattice does not have the Sperner property. 

\section{Stochastic Matrices}

We now investigate sets of stochastic matrices, with respect to which the following seminorm is always nonincreasing  
%As we mentioned, stochastic matrices leave nonincreasing the following seminorm  
$$\|x\|_\mathcal{P} = \frac{1}{2}(\max_i x_i - \min_i x_i).$$ 
The (polyhedral) unit ball of that seminorm\mods{:}
$$\mathcal{P} = \left\{x \; : \; \frac{1}{2}(\max_i x_i - \min_i x_i) \leq 1  \right\},$$
is thus invariant under multiplication by any stochastic matrix.

\begin{ex} \label{ex2} \modq{Suppose one wants to know whether all products made of the following two matrices converge.
$$
%\begin{pmatrix} 
%1&0&0 \\
%0&0&1 \\
%1&0&0
%\end{pmatrix}
A_1=\begin{pmatrix}
.5&0&.5 \\
1&0&0 \\
0&.5&.5
\end{pmatrix}, \;
A_2=\begin{pmatrix}
0&1&0\\
.5&0&.5\\
1&0&0
\end{pmatrix}.$$
Since the matrices are stochastic, the seminorm $\|.\|_{\mathcal{P}}$ is nonincreasing under multiplication by these matrices, as can be seen in Figure \ref{imP}. }
\begin{figure}[h!]
\centering
\usetikzlibrary{calc}

%\begin{tikzpicture}[scale=1.2,looseness=1,auto,line width=.3mm,shorten >=3pt, x= {(-0.353cm,-0.353cm)}, z={(0cm,1cm)}, y={(1cm,0cm)}]

\begin{tikzpicture}[scale=1.3,looseness=1,auto,line width=.3mm, x= {(-.866cm,-.5cm)}, y={(.866cm,-.5cm)}, z={(0cm,1cm)}]

%\begin{tikzpicture}[scale=.6,looseness=1,auto,line width=.3mm,shorten >=3pt, x= {(-0.6095cm,-0.4265cm)}, z={(0cm,1cm)}, y={(0.933cm,-0.25cm)}]
%\begin{tikzpicture}[scale=1.2,looseness=1,auto,line width=.3mm,shorten >=3pt, x= {(-0.866cm,-0.5cm)}, z={(0cm,1cm)}, y={(0.866cm,-0.5cm)}]
\begin{scope}[every node/.style={font=\small\itshape}]
\tikzstyle{every node}=[draw=none];

% P
\coordinate (1) at (1,0,0);
\coordinate (2) at (1,1,0);
\coordinate (3) at (0,1,0);
\coordinate (4) at (0,1,1);
\coordinate (5) at (0,0,1);
\coordinate (6) at (1,0,1);

\coordinate (21) at (1,\decy,\decz);
\coordinate (22) at (1,1+\decy,\decz);
\coordinate (23) at (0,1+\decy,\decz);
\coordinate (24) at (0,1+\decy,\decz+1);
\coordinate (25) at (0,\decy,\decz+1);
\coordinate (26) at (1,\decy,\decz+1);

\coordinate (31) at (1,2*\decy,2*\decz);
\coordinate (32) at (1,1+2*\decy,2*\decz);
\coordinate (33) at (0,1+2*\decy,2*\decz);
\coordinate (34) at (0,1+2*\decy,2*\decz+1);
\coordinate (35) at (0,2*\decy,2*\decz+1);
\coordinate (36) at (1,2*\decy,2*\decz+1);

% repère
\coordinate (0) at (0,0,0);
\coordinate (0x) at (\sizerep,0,0);
\coordinate (0y) at (0,\sizerep,0);
\coordinate (0z) at (0,0,\sizerep);

% repère 2
\coordinate (20) at (0,\decy,\decz);
\coordinate (20x) at (\sizerep,\decy,\decz);
\coordinate (20y) at (0,\sizerep+\decy,\decz);
\coordinate (20z) at (0,\decy,\decz+\sizerep);

% repère
\coordinate (30) at (0,2*\decy,2*\decz);
\coordinate (30x) at (\sizerep,2*\decy,2*\decz);
\coordinate (30y) at (0,\sizerep+2*\decy,2*\decz);
\coordinate (30z) at (0,2*\decy,2*\decz+\sizerep);

% image par la matrice [.5 0 .5; 1 0 0; 0 .5 .5];
\coordinate (13) at (.5,1+\decy,\decz);
\coordinate (14) at (0,\decy,\decz+.5);
\coordinate (15) at (.5,\decy,\decz+.5);
\coordinate (16) at (.5,1+\decy,\decz+.5);
\coordinate (17) at (1,1+\decy,\decz+.5);
\coordinate (18) at (.5,\decy,\decz+1);

% image par la matrice [0 1 0; .5 0 .5; 1 0 0];
\coordinate (7) at (1,2*\decy,2*\decz);
\coordinate (8) at (0,.5+2*\decy,2*\decz);
\coordinate (9) at (0,.5+2*\decy,2*\decz+1);
\coordinate (10) at (1,.5+2*\decy,2*\decz);
\coordinate (11) at (1,.5+2*\decy,2*\decz+1);
\coordinate (12) at (0,1+2*\decy,2*\decz+1);

% dessin de P
\draw[-, line width=.3mm] (1) -- (2);
\draw[-, line width=.3mm] (2) -- (3);
\draw[-, line width=.3mm] (3) -- (4);
\draw[-, line width=.3mm] (4) -- (5);
\draw[-, line width=.3mm] (5) -- (6);
\draw[-, line width=.3mm] (6) -- (1);

\draw[-, line width=.3mm] (21) -- (22);
\draw[-, line width=.3mm] (22) -- (23);
\draw[-, line width=.3mm] (23) -- (24);
\draw[-, line width=.3mm] (24) -- (25);
\draw[-, line width=.3mm] (25) -- (26);
\draw[-, line width=.3mm] (26) -- (21);

\draw[-, line width=.3mm] (31) -- (32);
\draw[-, line width=.3mm] (32) -- (33);
\draw[-, line width=.3mm] (33) -- (34);
\draw[-, line width=.3mm] (34) -- (35);
\draw[-, line width=.3mm] (35) -- (36);
\draw[-, line width=.3mm] (36) -- (31);

% dessin du repère
\draw[->, line width=.3mm] (0) -- (0x);
\draw[->, line width=.3mm] (0) -- (0y);
\draw[->, line width=.3mm] (0) -- (0z);

% dessin du repère
\draw[->, line width=.3mm] (20) -- (20x);
\draw[->, line width=.3mm] (20) -- (20y);
\draw[->, line width=.3mm] (20) -- (20z);

% dessin du repère
\draw[->, line width=.3mm] (30) -- (30x);
\draw[->, line width=.3mm] (30) -- (30y);
\draw[->, line width=.3mm] (30) -- (30z);

% image par la matrice [0 1 0; .5 0 .5; 1 0 0];
\draw[dashed, -, line width=.3mm] (7) -- (10);
\draw[dashed, -, line width=.3mm] (10) -- (8);
\draw[dashed, -, line width=.3mm] (8) -- (12);
\draw[dashed, -, line width=.3mm] (12) -- (9);
\draw[dashed, -, line width=.3mm] (9) -- (11);
\draw[dashed, -, line width=.3mm] (11) -- (7);

% image par la matrice [.5 0 .5; 1 0 0; 0 .5 .5];
\draw[dashed, -, line width=.3mm] (13) -- (16);
\draw[dashed, -, line width=.3mm] (16) -- (14);
\draw[dashed, -, line width=.3mm] (14) -- (18);
\draw[dashed, -, line width=.3mm] (18) -- (15);
\draw[dashed, -, line width=.3mm] (15) -- (17);
\draw[dashed, -, line width=.3mm] (17) -- (13);

\end{scope}
\end{tikzpicture}
\caption{%The polyhedra $\mathcal{P}$, $A_1 \mathcal{P}$ and $A_2 \mathcal{P}$ are infinite in the direction $\textbf{1}$. Their cross-sections are represented (left) and of its image by matrices $A_1$ (center, dashed) and $A_2$ (right, dashed).
\mods{The cross-sections of polyhedra $\mathcal{P}$ (left), $A_1 \mathcal{P}$ (center, dashed) and $A_2 \mathcal{P}$ (right, dashed). The three polyhedra are infinite in the direction in the direction $\textbf{1}$.}}
\label{imP}
\end{figure}
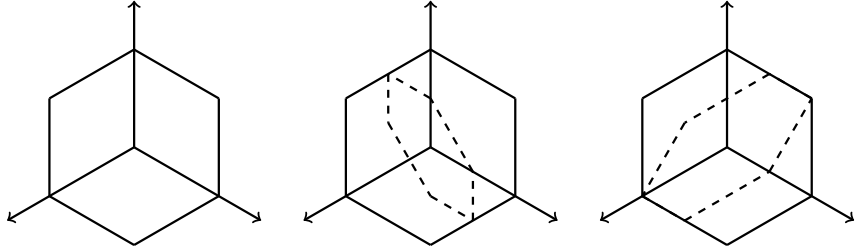
%\begin{figure}[h!]
%\centering
%\includegraphics[scale=.6]{Figure3-01.eps}
%\caption{The cross-sections of $\mathcal{P}$ (left) and of its image by matrices $A_1$ (center) and $A_2$ (right).}
%\label{imP}
%\end{figure}
\mods{
In this section, we will see (Theorem \ref{fin_cons}) that any infinite product of these two matrices converges to a rank one matrix if and only if any infinite periodic product, with period $\leq 3$  converges to a rank one matrix.
%One way to know whether all products of these two matrices converge, is to check all finite products up to a certain finite length. This length will be given in this section, by Theorem \ref{fin_cons} (it is equal to 3 in this case).
}
\end{ex}

We prove that the \modp{\mods{lattice of double-faces}} of this polyhedron has the Sperner property, allowing us to compute an explicit value for our bound $p^*$.

\begin{defi}[Upper and Lower Shadow \cite{Sperner_theory}] Let $(P, \preceq)$ be a graded poset and let $S \subseteq P$ be such that $\exists k, \; \forall x \in S, \;  \text{rank}(x) = k$.  We call the \emph{upper shadow}  $$\nabla(S) = \left\{x \in P : \exists y \in S, \; y \preceq x, \; \text{rank}(x) = k+1 \right\}.$$ %the set of elements such that . 
%Let $(P, \preceq)$ be a graded poset and let $S \subset P$ be such that $\exists k \text{ such that } \forall x \in S, \;  \text{rank}(x) = k$.  We call the \emph{upper shadow}  $$\nabla(S) = \left\{x \in P : \exists y \in S, \; y \preceq x, \; \text{rank}(x) = k+1 \right\}.$$ %the set of elements such that . 
Similarly, we define the lower shadow $$\Delta(S)= \left\{x \in P : \exists y \in S, \; x \preceq y, \; \text{rank}(x) = k-1 \right\}.$$
\end{defi}

\modq{We now describe the structure of the polyhedron $\mathcal{P}$: it} %$\mathcal{P}$ 
has no face of dimension 0 because $\forall x \in \mathcal{P}, a \in \mathbb{R}, \;  x+ a \textbf{1} \in \mathcal{P}$.
The face of dimension $n$ is equal to $\mathcal{P}$ itself.
\modp{Each \mods{double-face} of dimension $1 \leq d \leq n-1$ of $\mathcal{P}$ can be written as 
\begin{equation} \begin{aligned} F = \{x \in \mathcal{P} : & \forall i \in S_1, j \in S_2,  \; x_i = 2 - x_j\} \label{repr_face} \end{aligned} \end{equation}}
%Each face of dimension $1 \leq d \leq n-1$ of $\mathcal{P}$ can be written as 
%\begin{equation} \begin{aligned} F = \{x \in \partial\mathcal{P} : & \forall i \in S_m, \; x_i = \min_j x_j, \;  \\ & \forall i \in S_M, \; x_i = \max_j x_j\} \end{aligned} \end{equation}
for \mods{some} disjoint nonempty sets \modp{$S_1, S_2 \subset \{1, \dots, n\}$ with $|S_1 \cup S_2| = n-(d - 1)$.}

Therefore, the lower shadow of each single \modp{\mods{double-face}} $F$  of $\mathcal{P}$ of dimension $ 2 \leq d \leq n-1$ contains 
$$|\Delta(\{F\})| = 2(d-1)$$ elements (the \modp{\mods{double-faces}} obtained by adding an element to either \modp{$S_1$ or $S_2$}) and the upper shadow has
$$|\nabla(\{F\})| = n-d+1 \text{ or } |\nabla(\{F\})| = n-d$$ elements  (the \modp{\mods{double-faces}} obtained by removing an element from either \modp{$S_1$ or $S_2$}\mods{,} while keeping them both nonempty).

\begin{thm} The \modp{\mods{lattice of double-faces}} of $\mathcal{P}$ has the Sperner property. \mods{Its largest antichain is the set of double-faces of dimension $d^* =  \lfloor n/3 \rfloor + 1$.}
\begin{proof}Let $S = \{F_1, \dots F_{|S|}\}$ be any set of \modp{\mods{double-faces}} of $\mathcal{P}$ of the same dimension $d$, that is, a subset of a \mods{rank} level in the \modp{\mods{lattice of double-faces}}.
%Let $S$ be a set of faces of dimension $d$ and 
Let $E_+$ be the set of pairs of \mods{double-faces} of \modq{respectively} $S$ and $\nabla(S)$ being neighbors to each other: $$E_+ = \{(F_1, F_2) : F_1 \in S, \; F_2 \in \nabla(\{F_1\})\}.$$ %$F_1 \in S$, $F_2$
Since the upper shadow of each element of $S$ has at least $n-d$ elements%(all of them being in $\nabla S$ by definition)
, we have
$$|E_+| \geq |S| (n-d).$$
Since the lower shadow of each element of $\nabla S$ contains exactly $2d$ elements -- not all of which belonging to $S$ --, we have
$$|E_+| \leq |\nabla (S)| 2d.$$
Combining the two inequalities, we obtain \mods{ $|\nabla(S)| \geq |S| \frac{n-d}{2d}$ and
\begin{equation}   \forall  d \leq \frac{n}{3}, \;\; |\nabla(S)| \geq |S|. \label{up} \end{equation}
%\begin{align} |\nabla(S)| &\geq |S| \frac{n-d}{2d} \nonumber \\
%\forall  d \leq \frac{n}{3}, \;\; |\nabla(S)| &\geq |S| . \label{up} \end{align}
By a similar reasoning, we obtain $ |\Delta(S)| \geq |S| \frac{2(d-1)}{n-d+2}$ and
\begin{equation} \forall d \geq \frac{n+4}{3}, \;\; |\Delta(S)| \geq |S|. \label{down} \end{equation}}
%\begin{align} |\Delta(S)| &\geq |S| \frac{2(d-1)}{n-d+2} \nonumber \\
%\forall d \geq \frac{n+4}{3}, \;\; |\Delta(S)| &\geq |S|. \label{down} \end{align}
Let now $X$ be the largest antichain, let $d^-$ be the smallest dimension \modq{of an element in $X$}% in which $X$ contains elements 
and let $S^-$ be the intersection of the antichain with the level $d^-$. If $d^- \leq \frac{n}{3}$, Equation (\ref{up}) tells us that the antichain $$(X \backslash S^-) \cup \nabla(S^-)$$ has at least as many elements as $X$.
We can repeat this process until the antichain contains only faces of dimension strictly larger than $\frac{n}{3}$.  Similarly we use (\ref{down}) to obtain an antichain with at least as many elements of rank strictly smaller than $ \frac{n+4}{3}$. 
Since
$$\frac{n}{3}<d< \frac{n+4}{3}$$ has a unique integer solution $d^* = \lfloor n/3 \rfloor + 1,$ the final antichain contains only faces of dimension $d^*$.
\end{proof}
\label{th3}
\end{thm}

\subsection{\modq{A} New Finiteness Bound for Consensus}
\mods{By Theorem \ref{th3}, the largest antichain in the lattice of double-faces is  the set of all double-faces of dimension $d^* = \lfloor n/3 \rfloor + 1$.
From Equation (\ref{repr_face}), one can see that the number of double-faces of dimension $d$ is  $$f_{d} =  \binom{n}{d - 1} (2^{n - d} - 1)$$ }
%   $$f_{d} =  \binom{n}{d - 1} (2^{n - d+1} - 2).$$
%We have computed in \cite{CHJ12} that the number of faces of dimension $d$ of  $\mathcal{P}$  is 
% $$f_{d} = \binom{n}{d - 1} (2^{n - d+1} - 2).$$ 
\mods{and the size of the largest antichain is equal to}
%Therefore, for the case of stochastic matrices,
%Therefore, the answer to Question \ref{question} for the case of stochastic matrices is
\begin{equation}p^*  =  \binom{n}{\lfloor n/3 \rfloor} (2^{n- \lfloor n/3 \rfloor - 1} - 1). \label{p_cons}\end{equation} 
%\begin{equation}p^*  = \frac{f_{d^*}}{2} = \binom{n}{\lfloor n/3 \rfloor} (2^{n- \lfloor n/3 \rfloor - 1} - 1). \label{p_cons}\end{equation} %for \modif{seminorm $\|.\|_{\mathcal{P}}$.%for Polyhedron $\mathcal{P}$.}
%We now have an explicit value for the answer to Question \ref{question} in the case of stochastic matrix: 
%$$p^*  = \frac{f_{d^*}}{2} = \binom{n}{\lfloor n/3 \rfloor} (2^{n- \lfloor n/3 \rfloor - 1} - 1).$$

\mods{Combining this value of $p^*$ with Theorem \ref{main} and \cite[Proposition 1.a]{CHJ12} yields the next theorem.}
%\modq{Now that we have the exact value of the bound $p^*$, we can combine it with Theorem \ref{main} and \cite[Proposition 1.a]{CHJ12} to obtain the next theorem.}
%As we discussed in Section 1, contraction of the seminorm $\|.\|_{\mathcal{P}}$ is intimately linked with convergence. More formally, we have the following theorem.
\begin{thm} Let $\Sigma$ be a set of stochastic matrices. %System \eqref{sys} converges to consensus for any initial condition and any sequence of transition matrices from $\Sigma$ if and only if it converges to consensus for any initial condition and any \emph{periodic sequence}  of transition matrices from $\Sigma$, \emph{whose period is smaller than or equal to $p^* = \binom{n}{\lfloor n/3 \rfloor} (2^{n- \lfloor n/3 \rfloor - 1} - 1)$}.
Any left-infinite product of matrices from $\Sigma$ converges to a rank one matrix if and only if any \emph{periodic} left-\modq{infinite} product, \emph{with  period $ \leq p^* = \binom{n}{\lfloor n/3 \rfloor} (2^{n- \lfloor n/3 \rfloor - 1} - 1)$}, \modq{converges} to a rank one matrix.
%\begin{proof} This is a consequence  of Theorem \ref{main}, \moddd{of the value of $p^*$ \eqref{p_cons}} and of \cite[Proposition 1.a]{CHJ12}.
%\end{proof}
\label{fin_cons}
\end{thm}
%An infinite periodic product is simply the infinite power of a finite product. The convergence of powers of finite products  can be verified by checking that the modulus of the second\footnote{We recall that we investigate convergence to a rank one matrix and that the largest eigenvalue of every stochastic matrix is equal to one and corresponds to the eigenvector $\textbf{1}$.} eigenvalue of the product is smaller than 1 \cite{CHJ12}.
%The convergence  for all products with period bounded by $p^*$ can thus be verified by checking the second eigenvalue of every products of length smaller than or equal to $p^*$.

\modr{As announced in Example \ref{ex2}, if one wants to know if all infinite products of matrices from the set $\{A_1, A_2\}$ converge to a rank one matrix, 
Theorem \ref{fin_cons} implies that it is the case if and only if all infinite products with period\mods{s} $ \leq p^* = \binom{3}{\lfloor 3/3 \rfloor} (2^{3- \lfloor 3/3 \rfloor - 1} - 1) = 3$ converge  to a rank one matrix.}

A finiteness result such as Theorem \ref{fin_cons} was known \cite{bl-ol, Paz71} with $B = \frac{1}{2}(3^n - 2^{n + 1} + 1)$ instead of $p^*$. The new value $p^*$ is approximately equal to $\frac{3}{2 \sqrt{\pi n}} B$. %\frac{1}{2}(3^n - 2^{n + 1} + 1)$. 
Moreover, we prove next that Theorem \ref{fin_cons} is tight. 
%This is not a consequence of Theorem \ref{tightness_main}, 
%as the matrices constructed in the proof of Theorem \ref{tightness_main} are not necessarily stochastic.
%Theorem \ref{tightness_main} applied to polyhedron $\mathcal{P}$:
%For any dimension $n$, there is a set of matrices such that Theorem 1 is tight for $\mathcal{P}$.
%
%
\modq{
This is not a consequence of Theorem \ref{tightness_main}. Indeed, Theorem \ref{tightness_main} applied to polyhedron $\mathcal{P}$ guarantees that for any dimension $n$, there is a set of matrices  such that Theorem \ref{main} is tight for $\mathcal{P}$. However, the matrices in this set are not necessarily stochastic.} % Furthermore, in the proof of Theorem \ref{tightness_main}, the matrices that are constructed are not stochastic, e}

\begin{thm}
For any $n\geq 2$, there is a set of stochastic matrices such that:
\begin{itemize}
\item There is a product of length $p^*$ whose powers do not converge to a rank one matrix
\item For any product $P$ of length $\leq p^*-1$, the sequence of powers \modq{converges} to a rank one matrix.
\end{itemize}
\begin{proof}
 \mods{We will} construct \emph{stochastic} matrices that have the two properties:
\begin{equation}\forall i, \; A_i F_i \subseteq O_{(i \text{ mod } p^*) + 1} \label{prop1} \end{equation}
\begin{equation} \forall i, \; A_i (\mathcal{P} \backslash (F_i \cup - F_i)) 
\subseteq  \text{int}(\mathcal{P}). \label{prop2}\end{equation}

Then the same argument as in the proof of \modr{Theorem \ref{tightness_main}}  \mods{will allow} allow us to conclude. 
Recall that each face can be written as \begin{equation}F = \{x \in \partial\mathcal{P} : \forall i \in S_m, \; x_i = \min_j x_j, \;  \forall i \in S_M, \; x_i = \max_j x_j\}\end{equation}
%$$F = \{x \in \partial\mathcal{P} : \forall i \in S_m, \; x_i = \min_j x_j, \;  \forall i \in S_M, \; x_i = \max_j x_j\}$$
%\begin{equation} \begin{aligned} F = \{x \in \partial\mathcal{P} : & \forall i \in S_m, \; x_i = \min_j x_j, \;  \\ & \forall i \in S_M, \; x_i = \max_j x_j\} \end{aligned} \end{equation}
for certain disjoint nonempty sets $S_m, S_M \subset \{1, \dots, n\}$.
\modq{Let $F_i$ be a face \mods{such that} $S_{mi} = \{1, \dots, a_i\}$  and $S_{Mi} = \{n-c_i + 1, \dots, n\}$ for some $a_i$ and $c_i$ and similarly let $F_j = F_{(i \mod p^*) + 1}$ be such that $S_{mj} = \{1, \dots, a_j\}$  and $S_{Mj} = \{n - c_j + 1, \dots, n\}$ for some $a_i$ and $c_i$. Let $b_i = n - a_i - c_i$ and $b_j = n - a_j - c_j$.}
One matrix satisfying properties (\ref{prop1}) and (\ref{prop2}) is 
%$$A_i = \begin{pmatrix} +_{k_{1i} \times k_{1j}} & 0 & 0 \\
%+ & +_{(k_{2i}-k_{1i}) \times (k_{2j}-k_{1j})} & + \\
%0 & 0 & +_{(n-k_{2i}) \times (n-k_{2j})}
%\end{pmatrix}$$
\modq{$$A_i = \begin{pmatrix} +_{a_j \times a_i} & 0 & 0 \\
+_{b_j \times a_i} & +_{b_j \times b_i} & +_{b_j \times c_i} \\
0 & 0 & +_{c_j \times c_i}
\end{pmatrix}$$}
%\modq{$$A_i = \begin{pmatrix} +_{a_i \times a_j} & 0 & 0 \\
%+_{b_i \times a_j} & +_{b_i \times b_j} & +_{b_i \times c_j} \\
%0 & 0 & +_{c_i \times c_j}
%\end{pmatrix}$$}
%$$
%C=
%\left(
%\begin{array}{c|c|c}
%+ & 0 & 0 \\
%\hline
%+ & + & + \\
%\hline
%0 & 0 & + 
%\end{array}
%\right)
%$$
%
%$$\begin{pmatrix} +_{k_{1i} \times k_{1j}} & 0 & 0 \\
%+_{(k_{2i}-k_{1i}) \times k_{1j}} & +_{(k_{2i}-k_{1i}) \times (k_{2j}-k_{1j})} & +_{(k_{2i}-k_{1i}) \times (n-k_{2j})} \\
%0 & 0 & +_{(n-k_{2i}) \times (n-k_{2j})}
%\end{pmatrix}$$
where $+$ represents a positive element chosen such that the sum of the elements on each row sum to one. \modr{Let us see why property (\ref{prop1}) is satisfied. Let $x \in F_i$, we have that the first $a_j$ elements of $A_i x$ are weighted averages of the \mods{first} $a_i$  elements of $x$ and therefore they are equal to $\min_k x_k$. Similarly, the \mods{last $c_j$} elements of $A_i x$ are weighted averages of the \mods{last} $c_i$ elements of $x$ and therefore, they are equal to  $\max_k x_k$. The remaining elements are weighted averages of all elements of $x$ and therefore they are strictly smaller than $\max_k x_k$ and strictly larger than $\min_k x_k$. These three facts imply $A_i x \in O_j$ and since it is the case for any $x \in F_i$, property (\ref{prop1}) is satisfied. Property (\ref{prop2}) is proved in a similar manner.}

Without the assumption on the specific form of the faces $F_i$ and $F_j$, the matrix $A_i$ is the same up to some permutations of the rows and of the columns.

\end{proof}
\label{tightness_cons}
\end{thm}

%We show in Appendix B that for $n \geq 8$, $$p^* \leq \frac{3}{2 \sqrt{\pi n}} B,$$ so that our bound $p^*$ improves the previously known bound B y a multiplicative factor of $\frac{3}{2 \sqrt{\pi n}}$. %The next proposition  compares these two values.
%\begin{itemize}
% what does it bound
%\item If possible and relevant: what's the implication in terms of the complexity of some algorithm
%\end{itemize}

%\begin{prop} For stochastic matrices, for $n \geq 8$, our bound $p^*$ improves the previously known bound B y a multiplicative factor equal to $\frac{3}{2 \sqrt{\pi n}}$: $$p^* \leq \frac{3}{2 \sqrt{\pi n}} B.$$
%\label{improve}
%\end{prop}
%The proof can be found in appendix B. Figure \ref{bounds_comparison} shows the ratio between $p^*$ and $B$.
%Our bound is always smaller than the previously known bound. Moreover, for stochastic matrices, when $n \leq 8$, it is smaller than $\frac{3}{2 \sqrt{\pi n}}$ times the old bound as can be observed In Figure \ref{bounds_comparison}. %, we have represented the ratio between our bound and the previously known bound. This ratio is always smaller than one. For $n \leq 8$, our bound is smaller than $\frac{3}{2 \sqrt{\pi n}}$ times the old bound. 

%\begin{figure}[h!]
%    \centering
%    \includegraphics[scale = .5]{bounds_comparison.eps}
%    \caption{In blue: ratio between our bound $p^*$ and the previous bound $B$. This ratio is below one (red) and close to $\frac{3}{2 \sqrt{\pi n}}$ (green).}
%\label{bounds_comparison}
%\end{figure} 

\vspace{-.8cm}

\section*{Conclusion}

Deciding the asymptotic convergence of long matrix products has various
applications in engineering and computer science \modifq{\cite{DaLa92, blondelcassaignejungers08}}. In this
paper, \modifq{we have studied this problem for the case where the given set of matrices admits a nonincreasing polyhedral seminorm}, and one wonders whether all
long products of these matrices \mods{map} the state space onto points whose
seminorm is equal to zero (the so-called consensus problem is a
particular case of this setting).  We have significantly improved the available bound
by leveraging the combinatorial structure of (an abstraction of) the
dynamical system described by these matrices.

We see several further directions for our work: a major tool in our
analysis is Lemma 1, \mods{derived from Lagarias and Wang's work.  In \cite{lagarias-finiteness}, they} also provide a similar result when the invariant set is not a
polyhedron, but has a more involved algebraic structure (namely,
piecewise \modifq{analytic}). We believe that our analysis could be
further applied to \mods{piecewise analytic seminorms}, but it is not clear whether there would be
particular relevant applications in that setting.

\section*{References}

\bibliography{references}

\def\cprime{$'$} \newcommand{\noopsort}[1]{} \newcommand{\singleletter}[1]{#1}
  \def\cprime{$'$}
\begin{thebibliography}{10}
\expandafter\ifx\csname url\endcsname\relax
  \def\url#1{\texttt{#1}}\fi
\expandafter\ifx\csname urlprefix\endcsname\relax\def\urlprefix{URL }\fi
\expandafter\ifx\csname href\endcsname\relax
  \def\href#1#2{#2} \def\path#1{#1}\fi

\bibitem{lagarias-finiteness}
J.~C. Lagarias, Y.~Wang, The finiteness conjecture for the generalized spectral
  radius of a set of matrices, Linear Algebra and its Applications 214 (1995)
  17--42.

\bibitem{jungers_lncis1}
R.~M. Jungers, The joint spectral radius, theory and applications, in: Lecture
  Notes in Control and Information Sciences, Vol. 385, Springer, 2009.

\bibitem{Hernandez-Vargas}
E.~A. Hernandez-Vargas, R.~H. Middleton, P.~Colaneri, Optimal and mpc switching
  strategies for mitigating viral mutation and escape, IFAC World Congress,
  2011.

\bibitem{Shorten142}
R.~Shorten, F.~Wirth, D.~Leith, A positive systems model of tcp-like congestion
  control: asymptotic results, IEEE/ACM Transactions on Networking 14~(6)
  (2006) 616--629.

\bibitem{DaLa92}
I.~Daubechies, J.~C. Lagarias, Sets of matrices all infinite products of which
  converge, Linear Algebra and its Applications 161 (1992) 227--263.

\bibitem{blondelcassaignejungers08}
V.~D. Blondel, J.~Cassaigne, R.~M. Jungers, On the number of
  $\alpha$-power-free words for $2 < \alpha \leq 7/3$, Theoretical Computer
  Science 410 (2009) 2823--2833.

\bibitem{MoisionOrlitskySiegel01}
B.~E. Moision, A.~Orlitsky, P.~H. Siegel, On codes that avoid specified
  differences, IEEE Transactions on Information Theory 47 (2001) 433--442.

\bibitem{BlTi2}
V.~D. Blondel, J.~N. Tsitsiklis, The boundedness of all products of a pair of
  matrices is undecidable, Systems and Control Letters 41 (2000) 135--140.

\bibitem{Bamieh}
B.~Bamieh, M.~R. Jovanovic, P.~Mitra, S.~Patterson, Coherence in large-scale
  networks: dimension dependent limitations of local feedback, IEEE
  Transactions on Automatic Control 57~(9) (2012) 2235--2249.

\bibitem{Jad_tac}
A.~Jadbabie, J.~Lin, Coordination of groups of mobile autonomous agents using
  nearest neighbor rules, IEEE Transactions on Automatic Control 48~(6) (2003)
  988--1001.

\bibitem{bhot05}
V.~Blondel, J.~Hendrickx, A.~Olshevsky, J.~Tsitsiklis, Convergence in
  multiagent coordination, consensus, and flocking, in: Proceedings of the
  44$^{th}$ IEEE Conference on Decision and Control, 2005.

\bibitem{boyd_consensus}
L.~Xiao, S.~Boyd, S.-J. Kim, Distributed average consensus with
  least-mean-square deviation, Journal of Parallel and Distributed Computation
  67~(1) (2007) 33--46.

\bibitem{Olfati-Saber}
R.~Olfati-Saber, J.~S. Shamma, Consensus filters for sensor networks and
  distributed sensor fusion, in: Proceedings of the 44$^{th}$ IEEE Conference
  on Decision and Control.

\bibitem{bl-ol}
V.~D. Blondel, A.~Olshevsky, How to decide consensus? {A} combinatorial
  necessary and sufficient condition and a proof that consensus is decidable
  but {NP}-hard, SIAM Journal on Control and Optimization 52~(5) (2014) 2707 --
  2726.

\bibitem{Paz71}
A.~Paz, Introduction to {P}robabilistic {A}utomata, Academic Press, New York,
  1971.

\bibitem{Wolf63}
J.~Wolfowitz, Products of indecomposable, aperiodic, stochastic matrices,
  Proceedings of the American Mathematical Society 15 (1963) 733--736.

\bibitem{AnTi77}
J.~M. Anthonisse, H.~Tijms, Exponential convergence of products of stochastic
  matrices, Journal of Mathematical Analysis and Its Applications 598 (1977)
  360--364.

\bibitem{CHJ12}
P.-Y. Chevalier, J.~M. Hendrickx, R.~M. Jungers, Efficient algorithms for the
  consensus decision problem, To appear in SIAM Journal on Control and
  Optimization.

\bibitem{ziegler}
B.~Ziegler, Lectures on polytopes, in: Graduate Text in Mathematics,
  {Springer}, New York, 1995.

\bibitem{Sperner_theory}
K.~Engel, Sperner Theory, Cambridge University Press, 1997.

\end{thebibliography}

%\newpage
%\pagestyle{empty}
%\bibliography{references}{}
%\bibliographystyle{plain}
%
%\newpage
%\pagestyle{empty}

\end{document}